\documentclass[a4paper,11pt]{article}
\usepackage{pos}

\title{Are Neutron Stars Rich In H-dibaryons?}

\author*[a]{Jesper Leong}
\author[a]{Anthony W. Thomas}
\author[a]{Pierre A.~M.~Guichon}

\affiliation[a]{CSSM and ARC Centre of Exellence for Dark Matter Particle Physics, Department of Physics,\\
  University of Adelaide, South Australia 5005, Australia}
\emailAdd{jesper.leong@adelaide.edu.au}

\abstract{The possible existence of an H-dibaryon near the $\Lambda-\Lambda$ threshold has still not been decided experimentally. This raises the question of the potential effects on neutron stars if it does exist. We explore the consequences within the quark-meson coupling model, using the excluded volume formalism. While the H is abundant in heavy stars the maximum mass is only lowered slightly by its presence. }

\FullConference{The XVIth Quark Confinement and the Hadron Spectrum Conference (QCHSC24)\\
 19-24 August, 2024\\
 Cairns Convention Centre, Cairns, Queensland, Australia\\}

\begin{document}
\maketitle

\section{Introduction}
The H-dibaryon is a 6-quark, spinless boson with isospin zero and baryon number 2~\cite{Jaffe:1976yi}. Its composition is uuddss.  Theoretical studies of the H have concluded that the H, if it exists, is likely to have a mass around the $\Lambda-\Lambda$ threshold~\cite{Mulders:1982da,INOUE201228,NPLQCD:2012mex,Shanahan:2013yta}. The elusive H has so far evaded experimental detection 
with the current lower limit at $M_H>2224$ MeV~\cite{Takahashi:2001nm}. However, there is a possibility that the H could exist within the cores of neutron stars (NS). Here we examine this issue within the framework of the quark-meson coupling (QMC) model, starting with its effects on the equation of state (EoS) of dense matter in $\beta$-equilibrium.

Heavy NS have a mass of order $2$ M$_\odot$ or more, giving them an inner core density of around $0.8-1.0$ fm$^{-3}$~\cite{Antoniadis:2013pzd, Riley:2021pdl, Fonseca:2021wxt, NANOGrav:2017wvv, Romani:2022jhd}. At these densities the argument for hyperons in $\beta$-equilibrium is particularly strong. The $\Lambda$ is the first hyperon to appear, at just above $3$ $n_0$ in QMC~\cite{Motta:2020xsg, Leong:2023yma, Leong:2023lmw}. Since the mass of the H is thought to be near the $\Lambda-\Lambda$ threshold, then the H, should it exist, would be expected to occur shortly after the $\Lambda$. However, like the hyperons, the H should lower the predicted maximum mass of the NS. This effect is expected to be more severe in the case of the H, since it is a boson with zero momentum, but this is countered by the interaction of the H with the surrounding baryonic medium~\cite{Glendenning:1998wp, Faessler:1997jg}.

There has only been a handful of work modeling the appearance of the H in NS~\cite{Glendenning:1998wp,Tamagaki:1990mb, Faessler:1997jg, Wu:2024vvw}. The intermediate range forces involving the H are created by the scalar and vector mesons with the coupling strengths $g_{\sigma,H}$ and $g_{\rho, H}$. These are often fitted under the assumption that the H is present within the star, matching available NS observations~\cite{Glendenning:1998wp, Wu:2024vvw}. QMC is attractive in terms of modeling the H because the coupling strengths are predicted within the theory and there is no freedom to adjust $g_{\sigma,H}$ and $g_{\rho, H}$. Either the H exists and is capable of producing a heavy mass star, or it is incompatible. 

In the following section we outline the formalism of incorporating the H into the nuclear EoS. The theory used here is QMC using the excluded volume approach, with its details available in Refs.~\cite{Guichon:2018uew,Rikovska-Stone:2006gml, Leong:2023lmw}. The results are then presented with heavy NS constraints and the ramifications of these are discussed. We finish with a short conclusion on whether NS are really H-matter stars. 

\section{QMC Equation of State with the H-particle}
\begin{figure}
    \centering
    \includegraphics[width=0.7\linewidth]{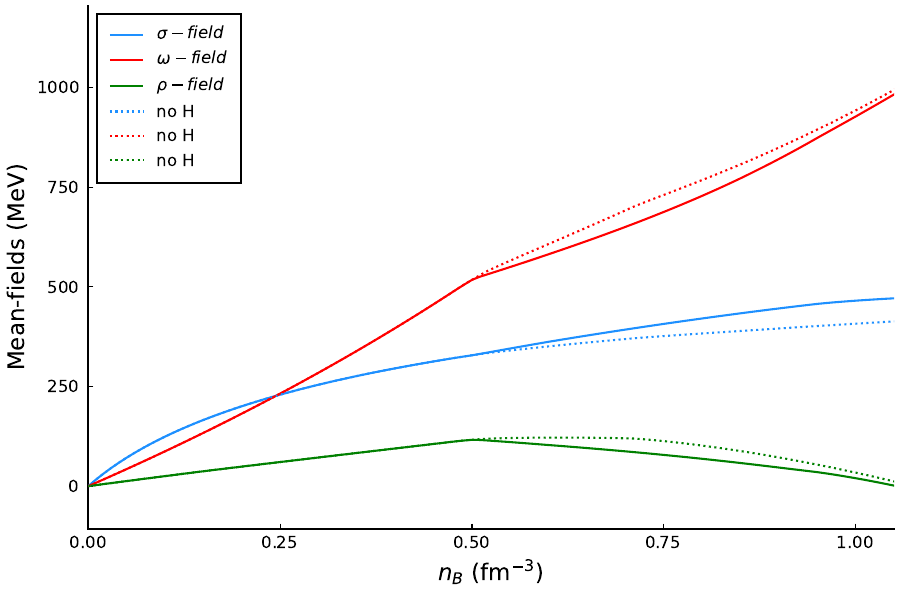}
    \caption{The mean-fields with neutron couplings are shown, with and without the H dibaryon. 
    $M_H=2247$ MeV is used here.}
    \label{fig:meanfields}
\end{figure}

The Lagragian for the H uses a minimal coupling scheme suggested by~\cite{Glendenning:1998wp}. 
\begin{equation}
\label{eq:Hlag}
    \mathcal{L}_D=\mathcal{D}^*_\mu H^* \mathcal{D}^\mu H - {M^*_H}^2 H^*H \, .
\end{equation}
Here $\mathcal{D}_\mu$ is defined as $\mathcal{D}_\mu=\partial_\mu+ig_{\omega,H} \omega_\mu$, which is the standard replacement for the vector $\omega$ field. $M_H^*$ is the effective mass and in QMC is given in equation~(\ref{eq:effMH}). The equation of motion for the H is presented in Ref.~\cite{Glendenning:1998wp}. Of immediate concern is the energy density and the mean-field contributions. Since the H does not carry isospin, only the $\sigma$ and $\omega$ field equations are changed. The modifications to the mean fields are given in Eqs.~(\ref{eq:sigmafield}) and (\ref{eq:omegafield}), as follows:
\begin{eqnarray}
    \label{eq:sigmafield}
    {m_\sigma}^2\sigma+\lambda_3 \frac{{g_\sigma}^3}{2}\sigma^2=\sum_f n^s_f-\frac{\partial M^*_H}{\partial \sigma} n_H, \\
    \label{eq:omegafield}
    {m_\omega}^2 \omega = \sum_f g_{\omega,f} n^v_f + g_{\omega,H} n_H \, .
\end{eqnarray}
The $\lambda_3$ coefficients adjust the strength of the self interaction term necessary to lower the incompressibility~\cite{Martinez:2018xep}. $g_{\omega,H}=\frac{4}{3}g_{\omega,N}$, the strength of the coupling of the H to the $\omega$ field, is set by simple counting of non-strange quarks. This is the same scheme used in Ref.~\cite{Glendenning:1998wp}. $n_f^s$ and $n_f^v$ are the scalar and vector densities, with the summation over $n$, $p$, $\Lambda$, $\Xi^0$ and $\Xi^-$. The $\Sigma$ and $\Delta$ have previously been shown not to appear as discussed in Refs.~\cite{Rikovska-Stone:2006gml, Whittenbury:2013wma}. The contribution to the energy density is
\begin{equation}
    \label{eq:energyH}
    \epsilon_H=2 {M_H^*}^2 H^*H= M^*_H n_H \, .
\end{equation}
We see that the H does not contribute directly to the pressure but there is an indirect contribution through the modification of the mean-fields described in Eqs.~(\ref{eq:sigmafield}) and (\ref{eq:omegafield}). 
These resulting changes to the mean-fields are displayed in Fig.~\ref{fig:meanfields}. 
\begin{eqnarray}
    M^*_H&=M_H-2\times[0.6672+0.0462 R_H-0.0021 {R_H}^2] g_\sigma \sigma \nonumber \\
    &+2\times[0.0016+0.0686R_H-0.0084 {R_H}^2](g_\sigma \sigma)^2 \, . 
    \label{eq:effMH}
\end{eqnarray}

The effective masses for the baryon octet in QMC are given in Ref.~\cite{Rikovska-Stone:2006gml}. To model $M^*_H$, we note that the H has twice the quark composition of the $\Lambda$. In QMC the mesons only couple to the $u$ and $d$ quarks.
Note that the scalar polarisability, $d$, encodes the strength of the 3-body force on the H. This is known to be sensitive to the free bag radius in the MIT bag model, $R_N^{free}$, which is used to confine the quarks. Because the H is a 6-quark state, $R_H^{free}$ is bigger than $R_N^{free}$. Mulders and Thomas found that $R_H^{free}=1.2 \times R_N^{free}$ \cite{Mulders:1982da}. Thus the effective mass of the H is expressed as above in Eq.~(\ref{eq:effMH}). We use $M_H=[ 2247, 2258, 2269 ]$ MeV as suggested in Ref.~\cite{Shanahan:2013yta}.

NS are taken to be at zero temperature and electrically neutral. The $\beta$-equilibrium condition must satisfy electrical charge and baryon number conservation. The H is electrically neutral but has a baryon number of 2. Thus, the total baryon number density is $n_B=\sum_f n_f +2n_H$, where $n_f$ is summed over the baryons and $n_H$ is the density of the H.

To model the short distance repulsion crucial to the NS EoS~\cite{Leong:2023yma, Aguirre:2002xr} acting on all baryons, including the H, we use the excluded volume correction~\cite{Rischke:1991ke} previously used to analyse the case with only nucleons and hyperons~\cite{Leong:2023lmw}. The EVE number density is $\Tilde{n}_f=\frac{n_f}{(1-v_0n_B)}$, with $n_B$ defined above. Thus, we interpreted this as H having twice the excluded volume, $v_0$, as the other baryons. For ease of comparison, we have chosen to use the same QMC parameters and a hard-core radius of $r=0.45$ fm as used in Ref.~\cite{Leong:2023lmw}, unless otherwise specified. 

\section{Results and Discussion}
\begin{figure}
    \centering
    \includegraphics[width=0.9\linewidth]{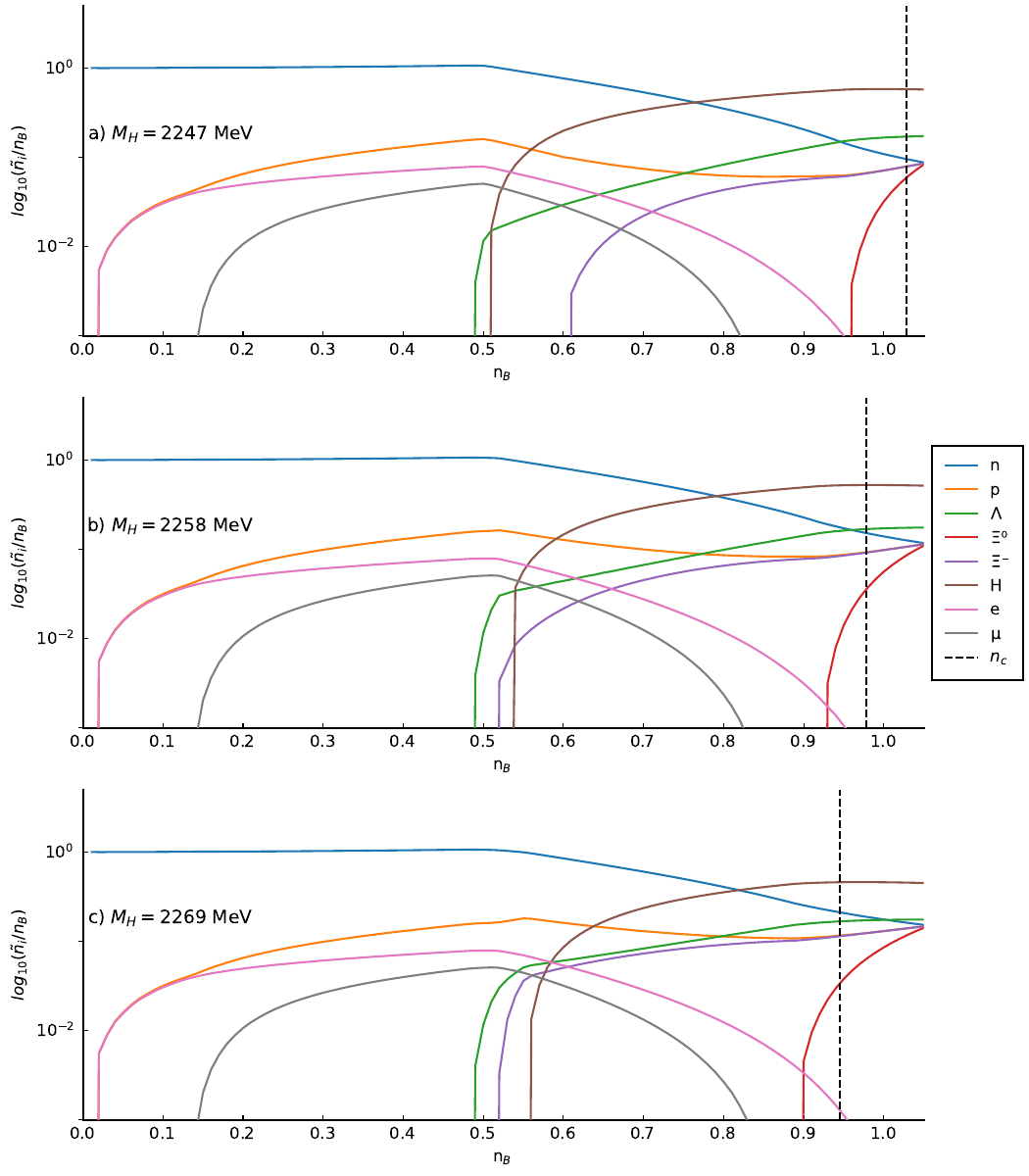}
    \caption{The EVE species fractions for a) $M_H=2247$ MeV, b) $M_H=2258$ MeV, and c) $M_H=2269$ MeV. For each $M_H$ the black dashed lines indicate the central number density, $n_c$, of the maximum mass NS.}
    \label{fig:species}
\end{figure}
Fig.~\ref{fig:species} displays how the H particle affects the NS composition. We first note that since the chosen $M_H>2M_\Lambda$, the H never appears before the $\Lambda$ hyperon. The larger $M_H$, the larger the density when it appears. The appearance of the H increases the density at which the $\Xi^0$ appears, and to a lesser extent the $\Xi^-$ for $M_H=2247$ MeV. How much the thresholds of the $\Xi^{0,-}$ increases is dependent on how soon the the H-dibaryon appears. The vertical black lines indicate the central density of the maximum mass neutron star for the given mass. For larger $M_H$ values, the central density is actually smaller. The H does alter the abundances of the nucleons and the hyperons and near maximum the H is the most populous particle within the cores of heavy NS.
\begin{figure}
    \centering
    \includegraphics[width=0.9\linewidth]{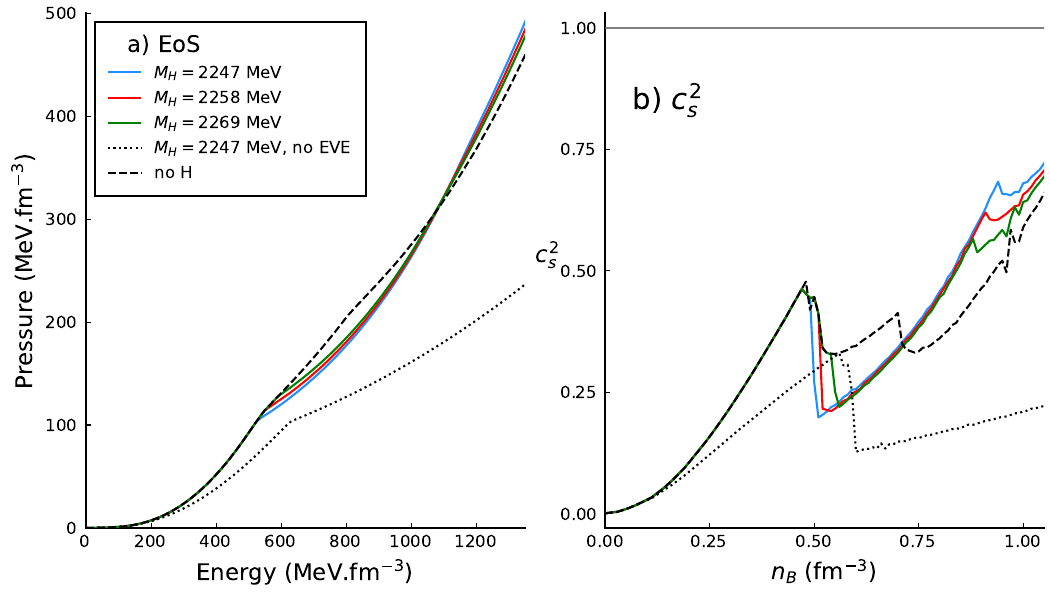}
    \caption{In a) we display the EoS with the accompanying b) speed of sound relative to light square, $c_s^2$. We have included the EoS without the H from Ref. \cite{Leong:2023lmw} as the black dashed line. The dotted black curve are the results for $M_H=2247$ MeV without the EVE.}
    \label{fig:EOSsound}
\end{figure}

In Fig.~\ref{fig:EOSsound} we show the results for a) the EoS and b) $c_s^2$, which is the speed of sound squared. We have added an additional EoS for $M_H=2247$ MeV, without the excluded volume correction, to show that the QMC coupling strengths (see Ref.~\cite{Leong:2023lmw} for refitting procedure) for $g_{\sigma,H}$ and $g_{\omega,H}$ can appropriately describe a star including an H-dibaryon which is stable against compression. The massive loss of pressure at around $0.6$ fm$^{-3}$ corresponds to the appearance of the H, and $c_s^2>0$ indicates that the values of $g_{\sigma,H}$ and $g_{\omega,H}$ used in the QMC model do not lead to an unstable star. 

It is known that the EVE may violate causality at higher densities. In Fig.~\ref{fig:EOSsound} b) we show that $c_s^2<1$ in all cases, which satisfies relativity. It is interesting to see that with the H, the EoS is actually stiffer at high density. This is explained by when the $\Xi^0$ hyperon first appears. When the H appears at around $0.5$ fm$^{-3}$ there is a significant loss of neutron degeneracy pressure. The same is true for hyperons. Without the H, the $\Xi^0$ appears at a little over $0.7$ fm$^{-3}$, as seen by the dashed line falling suddenly in Fig. \ref{fig:EOSsound} b). Recall that the appearance of the H actually increases the threshold for the $\Xi^0$, pushing it to a much higher density and not allowing it to soften the EoS. At this point the EVE exerted pressure has been allowed to build up and mitigate the degree in which the $\Xi^0$ softens the EoS. For larger values of $M_H$ compared to smaller $M_H$, the $\Xi^0$ appears slightly sooner, and the softening of the EoS can be readily viewed at around $0.9$ fm$^{-3}$ in fig. \ref{fig:EOSsound} b).

\begin{figure}
    \centering
    \includegraphics[width=0.7\linewidth]{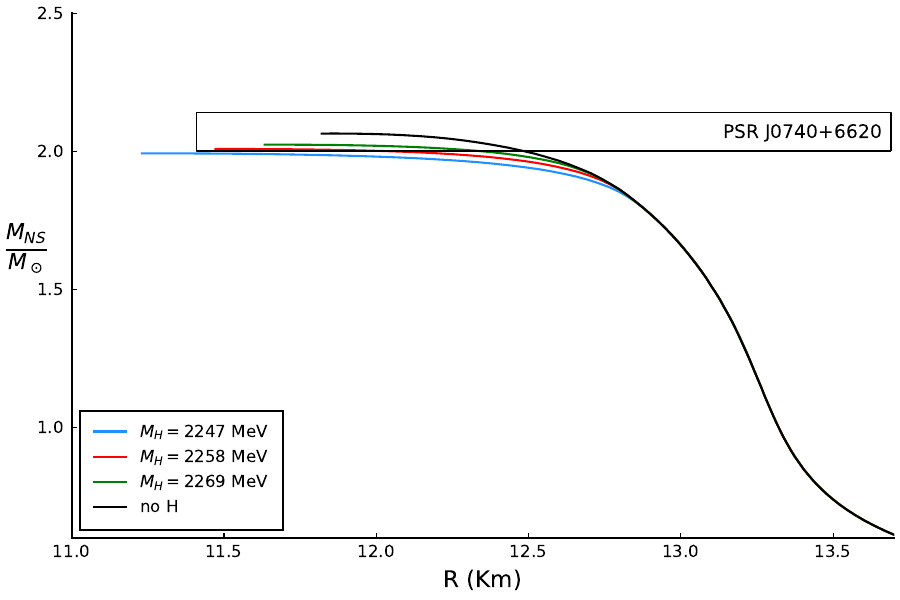}
    \caption{The TOV equation is used to produce the mass-radius curve for each of the EoS. The observed NS constrained used here is PSR J0740+6620~\cite{Riley:2021pdl, Fonseca:2021wxt}.}
    \label{fig:MRcurve}
\end{figure}

The stiffer EoS at high densities, when the H is included, does not, however, translate to heavier NS. In Fig.~\ref{fig:MRcurve} we show that the introduction of the H leads to a slight reduction of $M_{Max}^{NS}$. The closer $M_H$ lies to the $\Lambda-\Lambda$ threshold, the more the decrease in $M_{Max}^{NS}$. Once the H appears the  softening of the EoS leads to a mass plateau where the mass of the star hardly rises but becomes more and more compact. Note that the H does not appear for low mass stars and thus the tidal deformability is unaffected at $\Lambda_{1.4}=560$~\cite{LIGOScientific:2018cki}.

\section{Conclusion}
Our results do indicate that the H particle can be successfully incorporated into QMC with natural couplings to the $\sigma$ and $\omega$ fields. With $g_{\sigma,H}$ and $g_{\omega,H}$ fixed, the mass of the H determines the density at which the H appears at $\beta$-equilibrium. The H appears shortly after the $\Lambda$ hyperon and leads to only a slight reduction of the NS maximum mass. The H is highly abundant and in the heaviest of NS, it is predicted to be dominant, suggesting that the cores of these stars have a sizable H-matter condensate.

\bibliographystyle{elsarticle-num}
\bibliography{bibref}

\end{document}